\documentclass[aip,rsi, amsmath,amssymb, reprint]{revtex4-1}
\pdfoutput=1
\usepackage{graphicx}
\usepackage{dcolumn}
\usepackage{bm}
\usepackage{array}

\usepackage{pgfplots}
\usepackage{pgfplotstable}
\usepgfplotslibrary{external} 
\usepgfplotslibrary{fillbetween}
\usepgfplotslibrary{colorbrewer}
\pgfplotsset{compat=1.6}
\usepackage{lineno}

\usepackage{subfigure}
\usepackage{cleveref}

\pgfmathsetmacro{\Tfreeze}{364}
\pgfmathsetmacro{\Tjump}{442.335666891}
\pgfmathsetmacro{\tfreeze}{0.483233608052}
\pgfmathsetmacro{\tjump}{0.535762818744}
\pgfmathsetmacro{\Tc}{537}

\definecolor{android_blue}{RGB}{51,181,229}
\definecolor{android_dark_blue}{RGB}{0,153,204}
\definecolor{android_pink}{RGB}{170,102,204}
\definecolor{android_purple}{RGB}{156,39,176}
\definecolor{android_dark_pink}{RGB}{153,51,204}
\definecolor{android_green}{RGB}{153,204,0}
\definecolor{android_dark_green}{RGB}{102,153,0}
\definecolor{android_orange}{RGB}{255,152,0}
\definecolor{android_dark_orange}{RGB}{255,152,0}
\definecolor{android_red}{RGB}{255,68,68}
\definecolor{android_dark_red}{RGB}{204,0,0}
\definecolor{android_pink}{RGB}{156,39,176}
\definecolor{android_grey}{RGB}{158,158,158}

\pgfplotsset{grid style={dashed,grey,opacity=0.5}}

\pgfplotscreateplotcyclelist{peak_temp}{
{color=android_dark_blue,line width=1.0pt,mark=x,mark size=2pt,mark options={line width=1.0pt},line join=round},
{color=android_red,line width=1.0pt,mark=o,mark size=2pt,mark options={line width=1.0pt},line join=round},
{color=android_dark_green,line width=0.5pt,mark=|,mark size=2pt,mark options={line width=0.75pt},line join=round},
{color=black,line width=0.75pt,mark size=2pt,mark=triangle,mark options={line width=0.75pt},line join=round},
{color=android_blue,line width=0.5pt,mark=square,mark size=2pt,mark options={line width=0.75pt},line join=round},
	  {color=android_pink,line width=0.5pt,mark=diamond,mark size=2pt,mark options={line width=0.75pt},line join=round},
	  {color=android_orange,line width=0.5pt,mark size=2pt,mark options={line width=0.75pt},line join=round}
	  }

\begin{document}
\pgfplotsset{colormap/RdBu-9}


\title{Curie temperature modulated structure to improve the performance in heat-assisted magnetic recording  }

\author{O. Muthsam}
 \email{olivia.muthsam@univie.ac.at}
\author{C. Vogler}%
\author{D. Suess}
 \affiliation{ 
University of Vienna, Physics of Functional Materials, Boltzmanngasse 5, 1090 Vienna, Austria
}%

\date{\today}
             
\begin{abstract}
We investigate how a temperature reduction in $z-$direction influences the switching probability and the noise in heat-assisted magnetic recording (HAMR) for a bit in bit-patterned media with dimensions $d=5\,$nm and $h=10\,$nm. Pure hard magnetic bits are considered and simulations with a continuous laser pulse are performed using the atomistic simulation tool VAMPIRE. The results display that the switching behavior shows a thermally induced exchange spring effect. Simultaneously, both the AC and the DC noise increase. 
Additionally, we illustrate how an artificial Curie temperature gradient within the material can compensate the HAMR performance loss due to the temperature gradient. Further, due to the graded Curie temperature, DC noise can be reduced compared to a structure where no temperature gradient is considered.
\end{abstract}

\maketitle
             
\section{Introduction}
Heat-assisted magnetic recording (HAMR) is a technique to further increase the areal density (AD) in recording media in the future \cite{kobayashi,mee,rottmayer, kryder,thermomagnetic}. Conventional recording techniques are not able to overcome the so-called recording trilemma \cite{evans2}: To further increase the areal storage density of recording media, smaller grains are needed. These grains need to have a high anisotropy to be thermally stable. To write these high anisotropy grains, higher head fields are needed, which cannot be provided by a conventional write head. Thus, a heat pulse is included in the HAMR process. The temperature of the medium is locally enhanced which leads to a significant reduction of the coercive field and thus a reduction of the needed write field.\\
In the almost 60 years since HAMR was proposed \cite{burns}, quite a number of investigations have been performed to study HAMR in all details. However, to the best of our knowledge, most of these theoretical works do not take into account the temperature reduction in z-direction within the material in the writing process. Since the temperature of the heat pulse acts differently on different layers in the material in reality, it is interesting to study how this effects the performance of HAMR. We investigated how a temperature gradient in z-direction within the material influences the switching probability and the noise of a grain with a diameter of 5\,nm and a height of 10\,nm. To do this, pure FePt$-$like hard magnetic bits were considered and different temperature variations of the heat pulse in $z-$direction were assumed. Furthermore, we studied if an artificial Curie temperature gradient within the material in $z-$direction produced by a cleverly designed multilayer structure leads to the same effect as the temperature gradient and thus, can compensate the performance loss due to the decreased temperature. The simulations were performed with the atomistic simulation program VAMPIRE \cite{evans}, which solves the stochastic Landau-Lifshitz-Gilbert equation.
This work is structured as follows: In Section II, the recording assumptions as well as the considered material are introduced. The results are summarized in Section III and discussed in Section IV.

\section{Modeling HAMR}
A cylindrical recording grain is considered in the simulations with a height of 10\,nm and a diameter of 5\,nm. It can be interpreted as one grain of a state of the art heat assisted magnetic recording medium or one island of a patterned media design for ultra high density. In the atomistic simulations, only nearest neighbor exchange interactions between the atoms are included and a simple cubic crystal structure is used. In all simulations a continuous laser pulse with Gaussian shape and full width at half maximum (FWHM) of 20\,nm is assumed. The temperature profile of the heat pulse is given by

\begin{align}
T(x,y,t)= (T_{\mathrm{write}}-T_{\mathrm{min}})e^{-\frac{(x-vt)^2+y^2}{2\sigma^2}} + T_{\mathrm{min}} \\
= T_{\mathrm{peak}}\cdot e^{-\frac{(x-vt)^2}{2\sigma^2}} + T_{\mathrm{min}}
\label{pulse}
\end{align}
with
\begin{align}
\sigma=\frac{\mathrm{FWHM}}{\sqrt{8\ln(2)}}.
\end{align}

The speed $v$ of the write head is assumed to be 20\,m/s. $x_0=vt$ labels the down-track position of the write head with respect to the center of the bit. $x$ and $y$ denote the down-track and the off-track position of the grain, respectively. In our simulations both the down-track position $x$ and the off-track position $y$ are variable. The initial and final temperature of all simulations is $T_{\mathrm{min}}=270$\,K. In the simulations, a higher thermal gradient than currently available is used. This choice is justified by the fact that a higher thermal gradient will be needed to further increase the areal density in the future.
To simulate the reduction of the temperature within the material in $z-$direction, a different peak temperature $T_{\mathrm{peak}}$ is considered for every material layer. 
Initially, in each simulation, the magnetization of the investigated grain points in $+z$-direction. A continuous laser spot is moved over the grain. The applied field is modeled as a trapezoidal field with a write frequency of 1\,Ghz and a field rise and decay time of 0.1\,ns. The field strength is assumed to be 0.8\,T in $z$-direction. The trapezoidal field tries to switch the magnetization of the grain from $+z$-direction to $-z$-direction. At the end of every simulation, it is evaluated if the bit has switched or not.

\subsection{VARIATION OF T$_{\mathrm{\textbf{peak}}}$}
\begin{figure*}
\centering
\includegraphics[width=1.\linewidth]{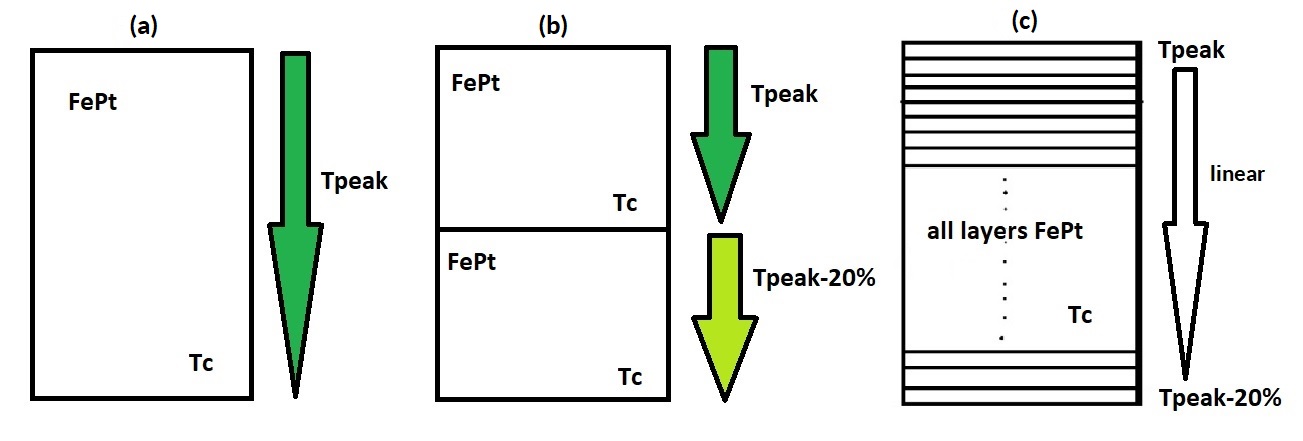}
\caption{Schematic representation of the considered hard magnetic structures. (a) The applied temperature is assumed to be constant within the material. (b) Two layers FePt with different peak temperatures of the applied heat pulse in the layers are assumed. (c) A material composition with 20 layers FePt and linear decreasing peak temperature in $z-$direction is considered.}
\label{allstructures}
\end{figure*}

A FePt like hard magnetic grain is considered and different temperature variations in $z$-direction are studied. The first "ideal" structure is pure FePt where no temperature gradient is assumed but a constant temperature applied to the material is considered. This material configuration can be seen in \Cref{allstructures}(a) and was already studied in former works. \cite{fundamental, areal, noisehamr} 
Secondly, a grain consisting of two identical layers FePt is considered where the only difference is the peak temperature of the applied heat pulse. Hereby, the peak temperature of the heat pulse applied to the top layer is assumed to be $20\%$ larger than that applied to the bottom layer (see \Cref{allstructures}(b)).
The last structure is a grain consisting of 20 identical layers FePt, see \Cref{allstructures}(c). Again, the difference between the applied peak temperature to the top layer and the applied peak temperature to the bottom layer is 20$\%$. In between, the peak temperature decreases linearly from top to bottom. Henceforth, the first material is called "(FePt)$_{N=1}$", while the second and third structures are named "(FePt)$_{N=2}$" and "(FePt)$_{N=20}$", respectively. 
The material parameters that are used for all structures can be seen in \Cref{tablemat}.

\begin{center}
\begin{table*}
\centering
\begin{tabular}{|>{\centering}m{1cm}|>{\centering}m{2cm}|>{\centering}m{2cm}|>{\centering}m{2cm}|>{\centering}m{2cm}|c|}\hline
     &$K_{1}$ (J/m$^3$)&$J_{ij}$ (J/link)&$\mu_{\mathrm{HM}}$  ($\mu_{\mathrm{B}})$&$J_{\mathrm{s}}$ (T)&$\alpha$\\
    \hline
	HM & $6.6\times 10^6$ & $5.18 \times 10^{-21}$ & 1.7& $1.42$ & 0.1\\
    \hline
 \end{tabular}
\caption{Material parameters of a FePt-like hard magnetic material. $K_{1}$ is the anisotropy constant. $J_{ij}$ is the exchange interaction within the material. The atomistic spin moment $\mu_{\mathrm{HM}}$ corresponds to the saturation magnetization $J_{\mathrm{s}}$. $\alpha$ is the damping constant.  }
\label{tablemat}
\end{table*}
\end{center}

\subsection{VARIATION OF T$_{\mathrm{\textbf{C}}}$}
Additional simulations are performed, where the Curie temperature of different layers is varied. The questions is, if a T$_{\mathrm{C}}$ gradient in z-direction with a spatially homogeneous applied temperature leads to the same results as an applied temperature gradient. 


The treated structures are similar to those considered above. First, a grain consisting of two layers FePt is considered where the Curie temperature of the bottom layer is $20\%$ larger than that of the bottom layer. This is done by increasing the exchange constant in the bottom layer by 20$\%$, since it holds 

\begin{align}
J_{ij}= \frac{3 k_{\mathrm{B}} T_{\mathrm{C}}}{\epsilon z}.
\end{align}

All other material parameters are taken as before.
The second structure is a grain consisting of 20 layers FePt where the difference between the Curie temperature of the top layer and that of the bottom layer is 20$\%$ with a linear increase from top to bottom. The first structure is denoted "(T$_{\mathrm{C}}$)$_{N=2}$" and the second one "(T$_{\mathrm{C}}$)$_{N=20}$".\\
In the course of the work, a temperature gradient is combined with a T$_{\mathrm{C}}$ gradient, to see if both gradients can compensate each other. Here, other than previously, the Curie temperature is decreased from top to bottom. The results are then compared with pure hard magnetic material. Again, a structure with two layers, named "(T$_{\mathrm{peak}}$+T$_{\mathrm{C}}$)$_{N=2}$", and one with 20 layers, called "(T$_{\mathrm{peak}}$+T$_{\mathrm{C}}$)$_{N=20}$", are considered.

\section{RESULTS}

\subsection{T$_{\mathrm{\textbf{peak}}}$ GRADIENT}

\begin{figure} 
\centering
\includegraphics[width=1.0\linewidth]{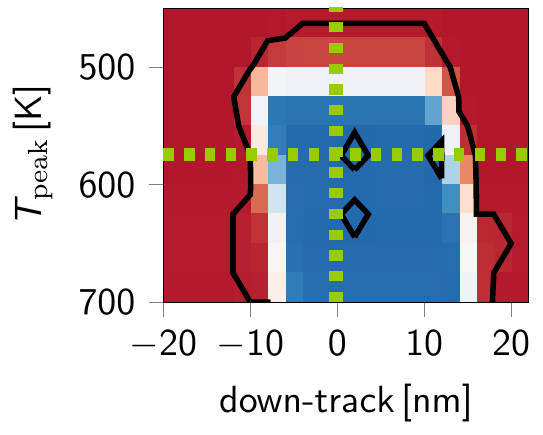}
 \caption{Switching probability phase diagram of a pure FePt like hard magnetic grain with constant temperature applied. The contour lines indicate the transition between areas with switching probability less than 1$\%$ (red) and areas with switching probability higher than 99.2$\%$ (blue). The dashed lines mark the switching probability curves in down-track and off-track direction that can be seen in \Cref{jitter}.}
  \label{feptphase}
\end{figure}

\begin{figure}
\centering
\subfigure{\includegraphics[width=1.0\linewidth]{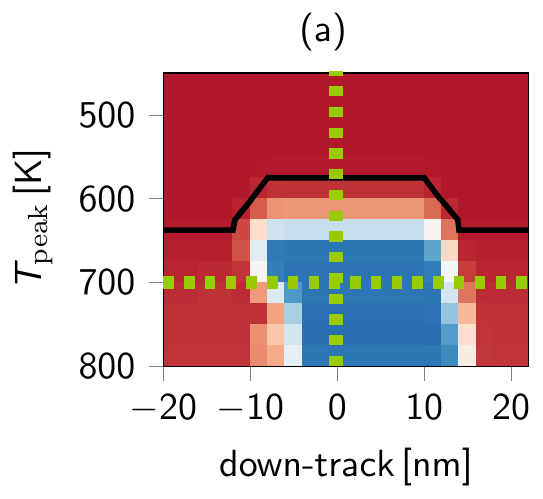}}
\subfigure{\includegraphics[width=1.0\linewidth]{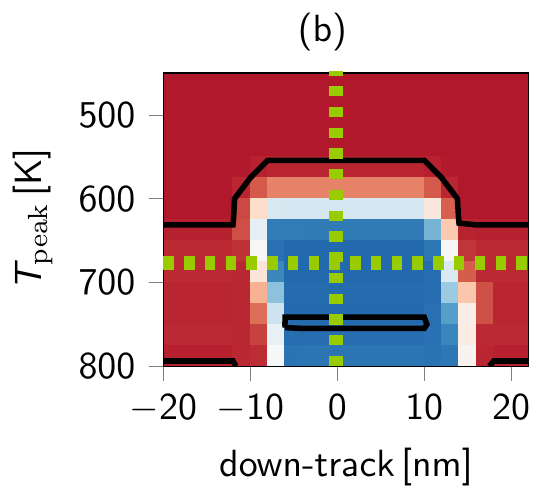}}
\caption{Switching probability phase diagram of a bit with (a) two layers FePt and (b) 20 layers FePt and an applied temperature gradient. The contour lines mark the transition between areas with switching probability less than 1$\%$ (red) and areas with switching probability higher than 99.2$\%$ (blue). The dashed lines mark the switching probability curves in off-track and down-track direction that can be seen in more detail in \Cref{jitter}.}
\label{phasenportraits}
\end{figure}

\begin{figure}
\centering
\subfigure{\includegraphics[width=1.0\linewidth]{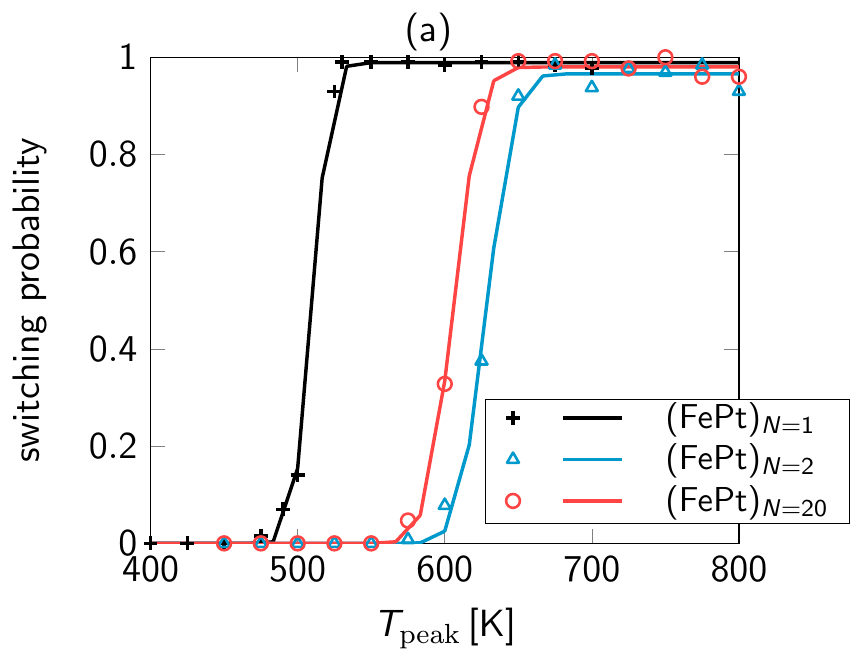}}
\subfigure{\includegraphics[width=1.0\linewidth]{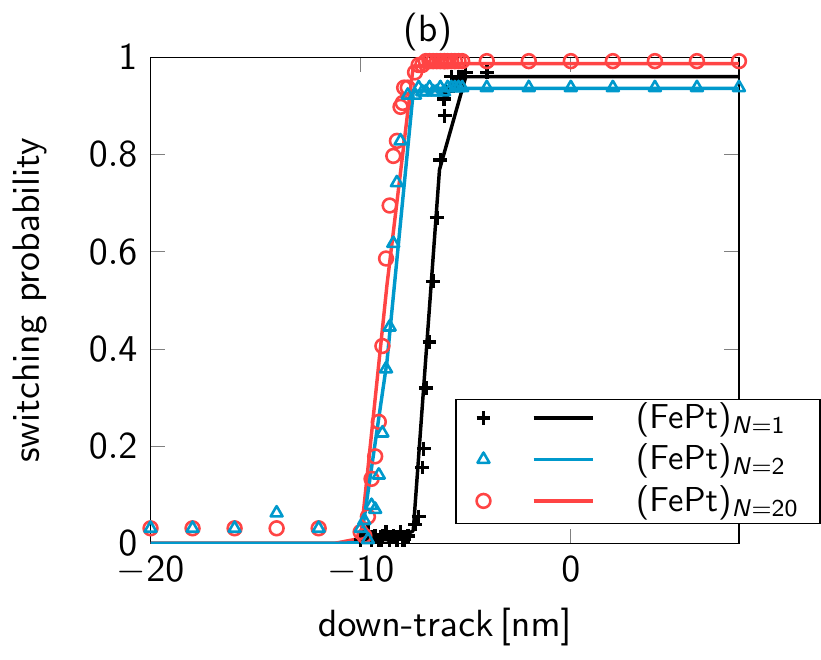}}
\caption{Switching probability curves of grains with applied temperature gradients marked by the dashed lines in \Cref{phasenportraits}. (a) shows $P(T_{\mathrm{peak}})$ corresponding to the down-track position $x=0$\,nm. In (b) $P(x)$ for a fixed off-track position $y=0$\,nm can be seen.}
\label{jitter}
\end{figure}

First, the switching probability of (FePt)$_{N=1}$ as a function of the down-track position $x$ and the off-track position $y$ is calculated and a phase diagram is determined. Solving
the equation 

\begin{align}
T_{\mathrm{peak}}=(T_{\mathrm{write}}-T_{\mathrm{min}})e^{-\frac{y^2}{2\sigma^2}}+T_{\mathrm{min}}.
\label{equation}
\end{align}

(see \cref{pulse}) for $y$ and fixing the write temperature $T_{\mathrm{write}}$, gives an equation where one can compute an unique off-track position $y$ for every peak temperature $T_{\mathrm{peak}}$. Thus, the phase diagram in \Cref{feptphase} shows the switching probability as a function of the down-track position $x$ and the, to $y$ corresponding, peak temperature $T_{\mathrm{peak}}$. The resolution in down-track direction is $\Delta x=$2\,nm and the resolution in temperature direction is $\Delta T_{\mathrm{peak}}=25$\,K. 
In each phase point 128 simulations are performed. Thus, the phase diagram contains about 60.000 switching trajectories, each with a length of 2\,ns. The vertical dashed line marks the down-track position $x=0\,$nm, whereas the horizontal dashed line indicates the off-track position $y=0$, if the peak temperature is $T_{\mathrm{peak}}=575\,$K.
Analogous to (FePt)$_{N=1}$, phase diagrams for (FePt)$_{N=2}$ and (FePt)$_{N=20}$ are determined. The results can be seen in \Cref{phasenportraits}(a) and (b). 
From the switching probability phase diagrams, one can see differences in the switching behavior of the three temperature gradients. (FePt)$_{N=1}$ only shows few areas of complete switching. Anyway, the phase diagram of (FePt)$_{N=2}$ shows no areas of complete switching at all. For (FePt)$_{N=20}$ the phase diagram shows 100$\%$ switching probability for only one temperature, namely $T_{\mathrm{peak}}=750\,$K. An interesting observation is that for (FePt)$_{N=1}$ switching starts at temperatures more than 20$\%$ lower than for the other temperature gradients.
To compare the switching behavior of the structures in more detail, the off-track jitter and the down-track jitter are computed. As mentioned before, the switching probability curves at down-track position $x=0\,$nm are marked by the vertical dashed lines in \Cref{feptphase} and \Cref{phasenportraits}. The resulting $P(T_{\mathrm{peak}})-$curves for temperatures in a range of $400\,$K and 800\,K with $\Delta T_{\mathrm{peak}}=25\,$K can be seen in \Cref{jitter}(a). A significant difference between the off-track probability curves of the different structures is the temperature that is needed to write the bit with a switching probability $>90\%$. Whereas the write temperature for pure (FePt)$_{N=1}$ is at approximately $T_{\mathrm{write}}=525\,$K (see \Cref{feptphase}, it is at $625\,$K for (FePt)$_{N=20}$ and $650\,$K for (FePt)$_{N=2}$, see the dashed horizontal lines in \Cref{phasenportraits}. In order to compare the jitter of the material configurations more accurately, the switching probability curves are fitted with a Gaussian cumulative distribution function 

\begin{align}
\Phi_{\mu,\sigma^2}=\frac{1}{2} (1 + \mathrm{erf}(\frac{x-\mu}{\sqrt{2\sigma^2}}))\cdot p
\label{distribution}
\end{align}

with

\begin{align}
\mathrm{erf}(x)=\frac{2}{\sqrt{\pi}} \int_0^x e^{-\tau^2} d\tau,
\label{error}
\end{align}

where the mean value $\mu$, the standard deviation $\sigma$ and the mean maximum switching probability $p \in [0,1]$ are the fitting parameters. The standard deviation $\sigma$, which determines the steepness of the transition function, is a measure for the transition jitter and thus for the achievable maximum areal grain density of a recording medium. The fitting parameter $p$ is a measure for the average switching probability for at the bit center. Fitting the $P(T_{\mathrm{peak}})$ curve of the material configurations gives parameters the jitter in temperature direction. These computed jitter parameters in temperature direction are then converted into an off-track jitter via the given thermal pulse profile and can be seen in \Cref{table}. 
Similar to the off-track jitter, the down-track jitter of the different structures is compared. The down-track switching probability curve at a certain off-track position can be determined by making a horizontal cut at the corresponding peak temperature in the phase diagram (see \Cref{feptphase} and \Cref{phasenportraits}). Since writing of the grain starts at higher temperatures for (FePt)$_{N=2}$ and (FePt)$_{N=20}$, the down-track jitter is calculated at different peak temperatures for the different compositions to make it comparable. The considered peak temperatures are approximately 10$\%$ higher than the minimal write temperature at which writing of the bit starts. Thus, the down-track switching probability curve is calculated at $T_{\mathrm{peak,FePt}_{N=1}}=575$\,K, $T_{\mathrm{peak,FePt}_{N=2}}=700$\,K and $T_{\mathrm{peak,FePt}_{N=20}}=675$\,K.  In \Cref{jitter}(b), the resulting $P(x)-$curves are shown. In comparison to the down-track jitter of (FePt)$_{N=1}$, the transitions of (FePt)$_{N=2}$ and (FePt)$_{N=20}$ are shifted to different down-track positions. Again, the switching probability curves are fitted with \cref{distribution} to determine the jitter parameter $\sigma$ for the different material configurations. The resulting down-track jitter parameters can be seen in \Cref{table}.

\begin{center}
\begin{table*}
\centering
\begin{tabular}{|l|c|c|c|c|>{\centering}m{1.5cm}|c|r|}\hline
   Material i &  $\sigma_{\mathrm{off}}$ & $\frac{\sigma_{\mathrm{off,i}}}{\sigma_{\mathrm{off,FePt}_{N=1}}}$ &$\sigma_{\mathrm{down}}$&$\frac{\sigma_{\mathrm{down,i}}}{\sigma_{\mathrm{down,FePt}_{N=1}}}$&$p_{\mathrm{off}}$&$p_{\mathrm{down}}$\\
  \hline
	(FePt)$_{N=1}$ & 0.326\,nm&---& 0.45\,nm&---&0.989&0.969\\
    \hline
    (FePt)$_{N=2}$ & 0.49\,nm& $+50\%$& 0.47\,nm&$+4.4\%$&0.988&0.936\\
 (FePt)$_{N=20}$ & 0.48\,nm & $+47.2\%$&0.52\,nm&$+15.5\%$&0.98&0.967\\
    \hline 
  (T$_{\mathrm{C}}$)$_{N=2}$ &0.87\,nm&$+166.8\%$&0.72\,nm&$+60\%$&0.966&0.968\\
  (T$_{\mathrm{C}}$)$_{N=20}$ &0.6\,nm&$+84\%$&0.6\,nm&$+33.3\%$&0.97&0.981\\
  \hline 
  (T$_{\mathrm{peak}}$+T$_{\mathrm{C}}$)$_{N=2}$&0.398\,nm&$+22\%$&0.44\,nm&$-2.2\%$&0.999&0.994\\
  (T$_{\mathrm{peak}}$+T$_{\mathrm{C}}$)$_{N=20}$&0.31\,nm&$-4.9\%$& 0.41\,nm&$-8.89\%$&0.998&0.996\\
  \hline
 \end{tabular}
\caption{Resulting jitter parameters of the simulations performed in this work. $p_{\mathrm{off}}$ and $p_{\mathrm{down}}$ are the fitting parameters $p$ for the switching probability curves in off-track and down-track direction, respectively.}
\label{table}
\end{table*}
\end{center}

\subsection{T$_{\mathrm{\textbf{C}}}$ GRADIENT}

\begin{figure}
\centering
\subfigure{\includegraphics[width=1.0\linewidth]{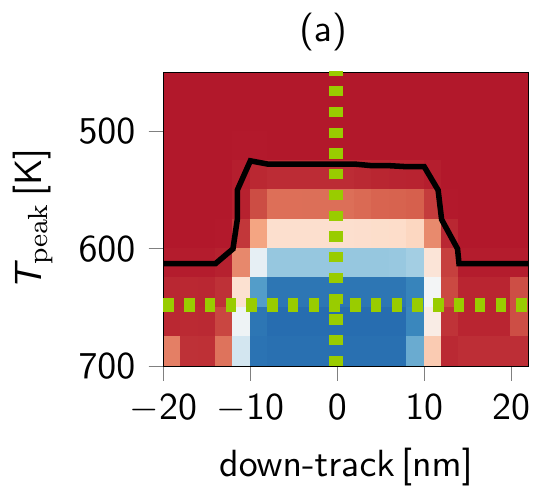}}
\subfigure{\includegraphics[width=1.0\linewidth]{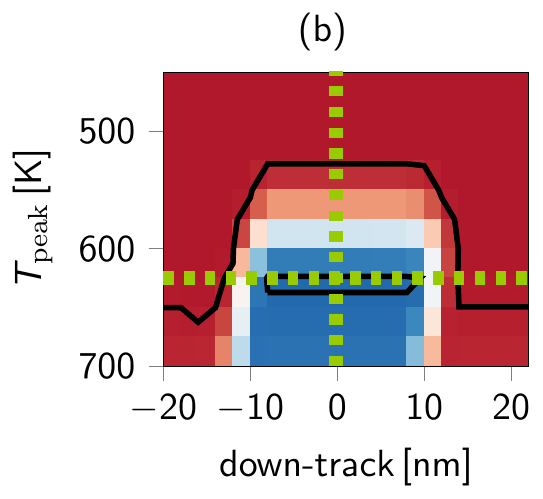}}
\caption{Switching probability phase diagram of a bit with (a) two layers FePt and (b) 20 layers FePt, when the Curie temperature is varied between the layers. The contour lines mark the transition between areas with switching probability less than 1$\%$ (red) and areas with switching probability higher than 99.2$\%$ (blue). The dashed lines mark the switching probability curves in off-track and down-track direction that are considered in more detail in \Cref{jittertc}.}
\label{phasenportraitstc}
\end{figure}

\begin{figure}
\centering
\subfigure{\includegraphics[width=1.0\linewidth]{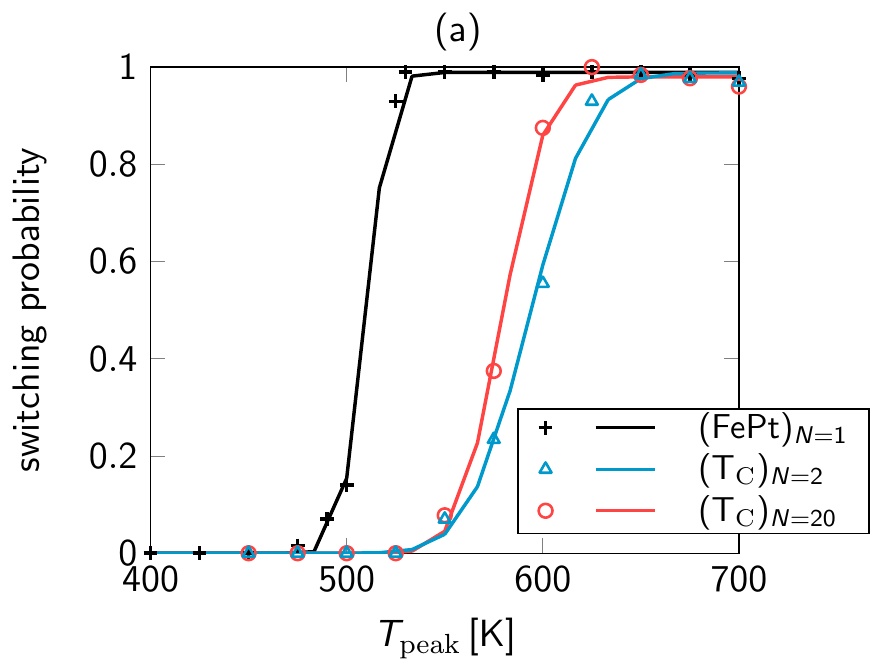}}
\subfigure{\includegraphics[width=1.0\linewidth]{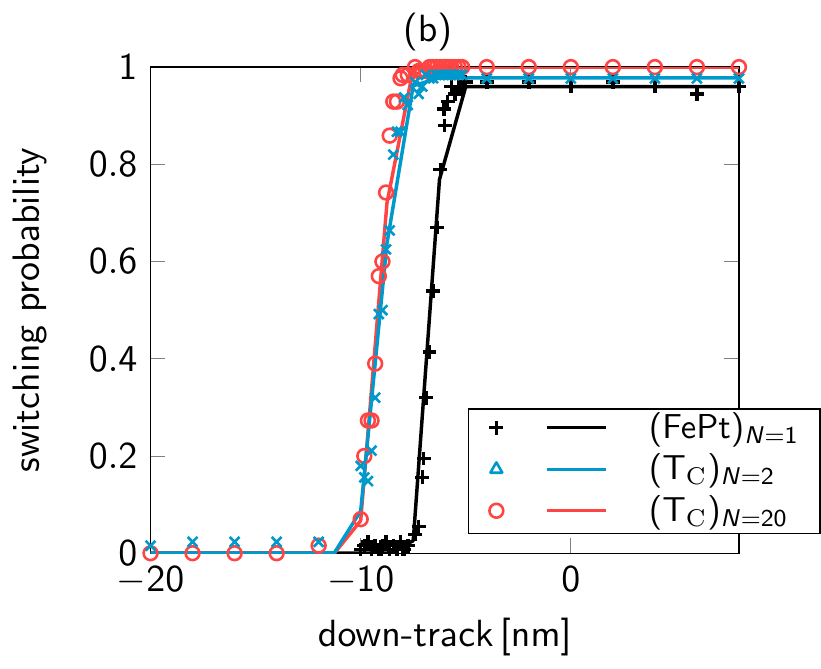}}
\caption{Switching probability curves of materials with T$_{\mathrm{C}}$ gradients at the positions marked in the phase diagrams in \Cref{phasenportraitstc}. In (a) one can see $P(T_{\mathrm{peak}})$ corresponding to the down-track position $x=0$\,nm. In (b) $P(x)$ for a fixed off-track position $y=0$\,nm is plotted.}
\label{jittertc}
\end{figure}

In this section, switching probability phase diagrams of the two considered structures with a T$_{\mathrm{C}}$ gradient are computed. The resulting phase diagrams can be seen in \Cref{phasenportraitstc}. The phase diagrams are similar to those of Section III.B., showing that both simulations qualitatively lead to the same results. However, there is one visible difference. One can see that switching starts at a temperature which is about 50\,K lower if a T$_{\mathrm{C}}$ gradient is considered instead of a temperature gradient. 
For better comparison, the switching probability curves in off-track and down-track direction are plotted in \Cref{jittertc}. The down-track switching probability curve of the different structures are again evaluated at different peak temperatures, namely $T_{\mathrm{peak,FePt}_{N=1}}=550$\,K, $T_{\mathrm{peak,T}_{C,N=2}}=650$\,K and $T_{\mathrm{peak,T}_{C,N=20}}=625$\,K. The switching probability curves are then fitted with a Gaussian distribution function. Both, the off-track and the down-track jitter parameters are summarized in \Cref{table}.
These fitting parameters show that compared to a temperature gradient both the off-track and the down-track jitter widen significantly in case of a T$_{\mathrm{C}}$ gradient. 

\subsection{COMBINATION T$_{\mathrm{peak}}$ - AND T$_{\mathrm{C}}$ GRADIENT}
\begin{figure}
\centering
\subfigure{\includegraphics[width=1.0\linewidth]{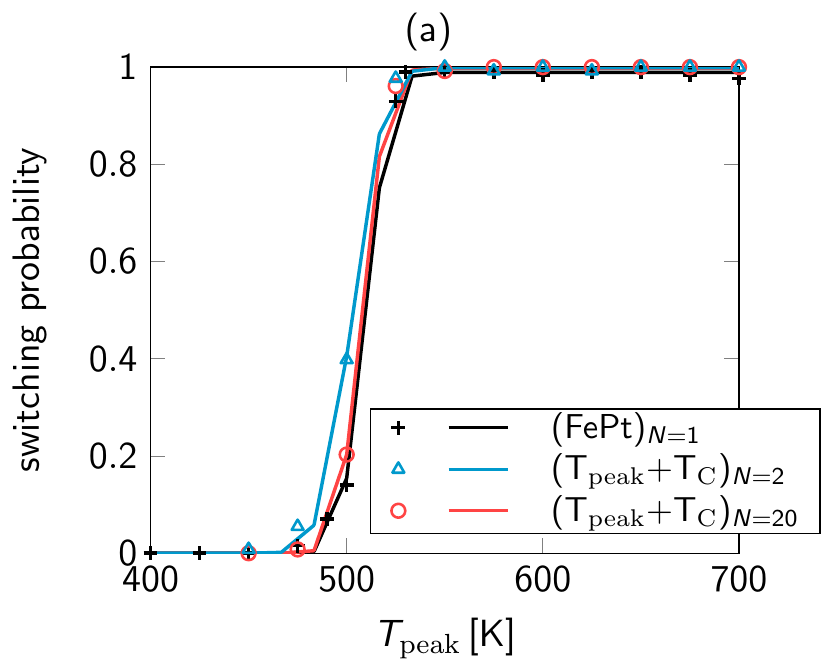}}
\subfigure{\includegraphics[width=1.0\linewidth]{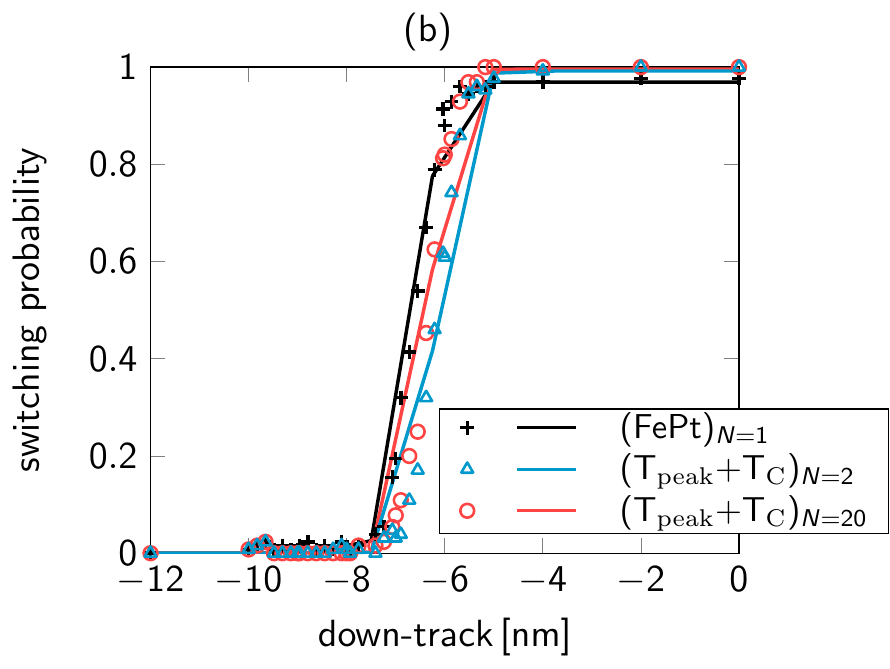}}
\caption{Switching probability curves of materials with combined $T_{\mathrm{peak}}$ and T$_{\mathrm{C}}$ gradients. The $P(T_{\mathrm{peak}})$ curve corresponding to the down-track position $x=0$\,nm is plotted in (a). In (b) one can see $P(x)$ for a fixed off-track position $y=0$\,nm.}
\label{jittertct}
\end{figure}

Finally, simulations are performed where a temperature gradient is combined with a T$_{\mathrm{C}}$ gradient. Here, only the switching probability curves in off-track and down-track direction are computed and then compared to those of pure hard magnetic material with a constant temperature applied. The resulting switching probability curves can be seen in \Cref{jittertct} (a) and (b). The switching probability curves in off-track direction of the combined structures look similar to that of (FePt)$_{N=1}$. An interesting difference is that the switching probabilities are higher for the combined structure than for (FePt)$_{N=1}$. More precisely, both combined structures reach 100$\%$ switching probability in the simulations for temperatures higher than 575\,K whereas (FePt)$_{N=1}$ only reaches complete switching for single temperatures. The switching probability curves in down-track direction are computed at the same peak temperature for all three structures, namely at T$_{\mathrm{peak}}=700\,$K.  Again, it can be seen that a combination of a temperature and a T$_{\mathrm{C}}$ gradient leads to higher switching probabilities than pure FePt.
Once more, the off-track and down-track jitter parameters are calculated via fitting of the curves. The off-track and down-track jitter parameters for the structures are visible in \Cref{table}. In fact, for (T$_{\mathrm{peak}}$+T$_{\mathrm{C}}$)$_{N=20}$ the off-track jitter even decreases slightly.\\ 
The down-track jitter parameters for the combined structures are even smaller than that of FePt$_{N=1}$. The higher switching probabilities can be clearly seen with the help of the fitting parameter $p$ in \Cref{table}, which is the mean maximum switching probability. 

\section{Discussion}
We investigated how a reduction of the temperature in $z-$direction within the material influences the switching probability of a cylindrical recording grain with a diameter of $5\,$nm and a height of $10\,$nm. Different hard magnetic material compositions were considered where the temperature in the different layers was either reduced in one step or linearly in 20 steps. Additionally, simulations with a T$_{\mathrm{C}}$ gradient within the material were studied. As anticipated, the switching probability reduces for hard magnetic structures when the temperature is reduced in $z-$direction. The results show, that for pure hard magnetic grains the bit switches at higher peak temperature and both the off-track and the down-track jitter increase if a temperature reduction in $z-$direction is assumed. In fact, the peak temperature that is needed to write the grain gets approximately 20$\%$ higher if the difference between the peak temperatures of the heat pulse in the top and the bottom layer is 20$\%$. This result is especially interesting for the case where the temperature is linearly decreased. Here, one could think that the magnetic moments of the layers with higher temperature, switch easier and induce the magnetic moments in the layers below to switch as well, just like a thermally induced exchange spring effect. It seems that only the weakest link in the system, i.e. the layer with the lowest applied peak temperature influences the temperature at which the grain is reliably switched.\\
The simulations with a Curie temperature gradient show the same qualitative behavior as those with a temperature gradient. However, a big difference is that switching starts at higher temperatures than for a T$_{\mathrm{peak}}$ gradient. Additionally, both the off-track and the down-track jitter are significantly larger than for the temperature gradient. The question remains, how much influence the increased jitter has on the SNR. Studies that will be published elsewhere, show that increasing the maximum switching probability from 0.98 to 1.0 leads to significant improvement of the SNR. Thus, this is an important factor to consider and shows that the maximum switching probability should be maximized. \\
As mentioned before, the simulations with the T$_{\mathrm{C}}$ gradient  and those with the temperature gradient qualitatively show the same behavior. In fact, the resulting difference could come from the different shapes of the applied heat pulse. For the Curie temperature gradient, a constant temperature is applied to the bit and thus the shape of the heat pulse is the same for every layer. In contrast to this, the peak temperature of the heat pulse is reduced in every layer for the T$_{\mathrm{peak}}$ gradient. Since the cooling time of the heat pulse stays the same in the simulation, the heat pulse applied to the top layer is steeper than that applied to the bottom layer. Eventually, this could result in the deviation between the two simulations.
However, since both variations show the same behavior, we combined a temperature gradient with a decreasing Curie temperature in the material and studied if both gradients compensate each other. The results show that both the off-track and the down-track jitter are similar to that of (FePt)$_{N=1}$ and even get smaller. Furthermore, a combination of a temperature and a T$_{\mathrm{C}}$ gradient leads to complete switching in a wider range of temperatures than for pure FePt with constant temperature. This is an very interesting result since one could use an exchange spring structure \cite{wangexchange,suessexchange,victora,suess, coffey, suess1} with a T$_{\mathrm{C}}$ gradient to compensate the HAMR performance loss due to the temperature gradient in $z-$direction with just varying the Curie temperature within the material.\\

\section{ACKNOWLEDGEMENTS}
The authors would like to thank the Vienna Science and Technology Fund (WWTF) under grant No. MA14-044, the Advanced Storage Technology Consortium (ASTC), and the Austrian Science Fund (FWF) under grant No. I2214-N20 for financial support. The computational results presented have been achieved using the Vienna Scientific Cluster (VSC).

\nocite{*}
\bibliographystyle{unsrt}
\bibliography{zdependence.bib}
\end{document}